\documentclass[a4paper,11pt]{article}
\DeclareUnicodeCharacter{2212}{\ensuremath{-}}
\usepackage{jheppub} 
\usepackage{lineno}
\usepackage{bm}

\arxivnumber{1234.56789} 

\title{Modeling Falling Backgrounds with Exponential Mixtures}

\author[1]{Austin Townsend,}
\author[1]{Marc Osherson,}
\author[1]{Mike Hildreth, }
\author[1]{and Stefano Castruccio}
\affiliation[1]{University of Notre Dame,\\
USA}

\emailAdd{atownse2@nd.edu}

\abstract{
Searches for new physics at the LHC often look for localized excesses on smoothly falling 
background distributions.
Several classes of background models have been considered, including polynomials and other 
parametric families; however, these approaches can require extensive analysis-specific 
development as datasets grow.
In this work, we motivate the finite exponential mixture as a flexible semi-parametric class 
of functions for approximating falling distributions, drawing on results from extreme value 
theory.
Using two published datasets ($n=28,619,185$ and $n=5,036$), we show that the exponential mixture performance is comparable 
to existing methods for both small and large datasets.
Finally, in simulation studies ($n = 5,036$), we find that the finite exponential mixture exhibits small bias relative to the true statistical uncertainty while maintaining consistent nominal coverage in the bulk.
}

\DeclareUnicodeCharacter{2009}{\,}
\begin{document}
\maketitle
\flushbottom

\section{Introduction}
\label{sec:intro}

Searching for new particles is a primary focus of high-energy physics, providing the main avenue for probing physics beyond the Standard Model. At the Large Hadron Collider (LHC), these searches can take the form of resonance searches—or "bump hunts"—where high-energy proton collisions are used to produce short-lived, novel states. If a new particle decays into visible final-state products within the detector, and a sufficient number of these events are recorded, its presence can be identified as a statistically significant localized excess, or "bump," in the reconstructed invariant mass. Standard Model (SM) processes can mimic these decays of new particles; however, in the regions of interest their contributions are often non-resonant. Isolating a genuine signal on top of these sometimes large SM contributions requires an accurate description of the SM expectation.

This paper focuses on background models for resonant searches, where signals are 
represented by localized (resonant) deviations on top of a smooth falling background. The signal 
shapes in non-local (non-resonant) searches vary and require background estimation methods, which 
are out of the scope of this paper.

Background models in resonance searches often rely on mass sidebands (regions 
of the spectrum surrounding a signal window) for identification, and range from 
very flexible to rigid, analysis-specific background models. Flexible models like polynomials are attractive for their universal approximation properties but may need a large number of parameters and can 
be hard to optimize. These models can also be too flexible, fitting the signal or generating nonphysical behaviors in the tails. Previous searches considering polynomials, 
such as those used to discover/measure the Higgs boson properties \cite{CMS_Higgs_Discovery, Aad_2014}, avoid these issues by only fitting a portion of the observed mass spectra near the resonance. Since the Higgs mass 
occurs within the bulk of the background, the background model is well constrained by the data 
sidebands; however, other analyses extend their search to the extrema of the distributions (tails), where no sidebands are 
available and better extrapolation techniques are required.

Alternatively, analysis-specific background models are often more rigid, avoiding some of the drawbacks that 
come with very flexible models like polynomials. The underlying theory describing the particle 
interactions is complicated enough that there are generally no closed-form expressions for 
describing these backgrounds, though there are some good approximations. A salient example is 
the ``dijet function'', first used in the 1990s by the CDF and D\O\ collaborations and more 
recently by CMS and ATLAS \cite{Harris:2011bh} to describe the invariant mass distribution of 
particle jet pairs in hadron colliders. The terms in this function are motivated by leading 
order quantum chromodynamics (QCD) and the mass dependence of parton distribution 
functions \cite{Harris:2011bh}, but more recently also contain empirically motivated terms. This function works on existing datasets, but 
since it relies on a number of approximations and ad hoc corrections, it is likely to need 
updating as datasets continue to grow in size.

The dijet function has also been applied in other contexts, such as analyses 
involving photons, despite the more complicated background composition 
in those channels. To combat the bias associated 
with the assumption of a particular functional form, more recent analyses \cite{CMS_DP1, 
CMS_high_mass_diphoton, CMS_DP2, CMS_DP3} will often consider a number of background functions 
simultaneously using the discrete profiling method \cite{discrete-profiling}. These functions 
are typically chosen from a larger set of function classes, which can be extended with more 
parameters. In these cases, researchers are typically interested in finding function classes that 
approximate the spectra with the smallest number of parameters. More broadly, there is significant interest in developing generic, 
reusable fitting solutions for resonance searches in high-energy physics, including 
Gaussian-process-based background models \cite{Frate_2017}, and automated symbolic regression \cite{SymbolFit_2025}.

In this paper, we consider the exponential mixture as an alternative class of functions for modeling 
smoothly falling backgrounds, motivated as a generalization of the unbounded asymptotic 
distribution of threshold excesses. Extreme Value Theory (EVT) provides a rigorous statistical 
framework for describing the behavior of the tails of distributions. Rather than modeling the 
full mass spectrum from first principles or relying on ad hoc functional forms, EVT allows us to 
justify our approach through asymptotic theory.

In addition to the mathematical motivations, we show that the finite exponential mixture can 
match the performance of previous methods in datasets that vary widely in size (from $n=5,036$ 
to $n=28,619,185$). Finally, in pseudo-datasets generated from alternative models, we show that, with an appropriate model 
selection metric, the finite exponential mixture has small bias and consistent coverage in the 
bulk. The exponential mixture is flexible enough to approximate motivated alternatives, yet sufficiently constrained that it cannot produce spurious peaks or other non-monotone structures.

The remainder of this paper is organized as follows. In Section~\ref{sec:emm} we introduce the 
finite exponential mixture and our parameterization. Section \ref{sec:examples} begins with the model selection procedure, implementation details, and global optimization. It is then divided into subsections covering two example datasets: the Run 2 ATLAS dijet dataset (\ref{sec:example1} and the Run 2 CMS high-mass diphoton dataset (\ref{sec:example2}). In 
Section~\ref{sec:toys} we generate pseudo-datasets using the functions from the Run 2 CMS high-mass diphoton analysis and compare the bias, coverage, and spurious signal of those models 
with the exponential mixture. Section~\ref{sec:discussion} discusses the results, extensions, and areas for future research. Finally the main results are summarized in Section \ref{sec:conclusion}.

\section{Exponential Mixtures}
\label{sec:emm}

A foundational result in EVT, the Pickands--Balkema--de Haan theorem
\cite{Picklands_1975, Balkema_Haan_1974} states that, beyond a sufficiently high threshold, the tails of a broad class of parent distributions converge to a Generalized Pareto Distribution 
(GPD). This framework is highly attractive for new physics searches where signal regions lie in the extreme kinematic tails past dominant Standard Model resonances. Additionally, it may apply to scenarios where relevant Standard Model processes have already been accounted for, allowing the GPD to isolate and model the remaining non-resonant component. The GPD density is given by:
\begin{equation}
f_{\text{GPD}}(x | u, \sigma, \xi) = \frac{1}{\sigma} \left( 1 + \xi \frac{x - u}{\sigma} 
\right)^{-\left( \frac{1}{\xi} + 1 \right)}
\end{equation}
where $u$ is the lower threshold, $\sigma$ is the scale parameter, and $\xi$ is the shape 
parameter. The GPD domains are strictly categorized by the value of $\xi$: for $\xi>0$, the 
distribution is heavy-tailed and unbounded; for $\xi=0$, it reduces to a simple exponential 
distribution; and for $\xi<0$, it terminates at a finite upper bound $x<u-\sigma/\xi$.

However, in practical applications, the strict assumptions of EVT may not hold across the entire 
range of interest. Asymptotic convergence can be slow, or the data may simultaneously span two 
distinct EVT regimes---such as an intermediate heavy-tailed behavior that eventually transitions 
into a bounded far-tail. A standard GPD is forced to commit to a single regime via $\xi$: 
choosing $\xi<0$ captures the upper bound but fails to model the heavy-tail behavior as the 
threshold $u$ is lowered, while choosing $\xi>0$ models the heavy tail but cannot accommodate a 
finite boundary. Furthermore, experimental constraints may force the selection of a lower 
threshold $u$ that sits close to unresolved resonance peaks, where these competing tail dynamics 
overlap. In these complex, pre-asymptotic, and mixed regions, standard GPD assumptions break 
down, requiring a more flexible modeling approach that can accommodate multiple regimes while 
remaining asymptotically well-behaved. Crucially, when $\xi>0$, the GPD can be mathematically 
framed as an exponential-gamma mixture also known as a Lomax or Pareto Type II distribution. This structural relationship directly motivates the use of more general exponential mixtures, which may provide the necessary flexibility to model 
pre-asymptotic and intermediate shapes.

Mixture models are a flexible class of models that can approximate a wide variety of 
distributions.
The density of a general single-parameter mixture can be written as:
\begin{equation}
\label{general_mixture}
f(x|\theta) = \int p(x | \theta) \pi(\theta) d\theta \, ,
\end{equation}
where $p(x|\theta)$ is the base density and $\pi(\theta)$ is the mixing density. In general, 
$\pi(\theta)$ can be any density function defined on the domain of $\theta$. Without assuming a specific functional form for $\pi(\theta)$, it is commonly estimated non-parametrically, yielding a discrete distribution represented by a linear combination of $k$ Dirac delta functions  \cite{Kiefer_Wolfowitz, Lindsay}.
This discrete estimate of $\pi(\theta)$ can be written as:
\begin{equation}
\label{sum_approx}
\pi_k(\theta|\vec\theta, \vec{w})=\sum_{i=1}^k w_i \delta(\theta-\theta_i),
\end{equation}
where $\theta_i$ represent the discrete support points (i.e., locations of the Dirac deltas), 
and the weights $w_i$ satisfy the normalization condition $\sum w_i=1$ (with $w_i\geq0$) to 
ensure a well-defined probability distribution. Substituting eq.~\eqref{sum_approx} into 
eq.~\eqref{general_mixture}
we find the common form of the finite mixture density:
\begin{equation}
\label{finite_mixture}
f_k(x|\vec\theta) = \sum_{i=1}^k w_i p(x | \theta_i) \, .
\end{equation}

The class of exponential mixtures is able to approximate any completely monotone function\footnote{\label{cm_def}A function $f(x)$ is completely monotone if $(-1)^n \frac{d^n}{dx^n}f(x) \geq 0$ for all $n \in \{0,1,2,\ldots\}$, meaning the function is non-negative, non-increasing, convex, etc.} arbitrarily closely \cite{bernstein}. The class of completely monotone functions is closed under addition and multiplication. For functions in this class, the finite exponential sum is guaranteed to converge uniformly \cite{McGlinn_1978}.
This class does not include functions like the falling tail of a Normal distribution, for 
example, though the approximation becomes better if one restricts the domain to be sufficiently far from the mean. In some regimes the Normal distribution tail converges to the GPD with 
$\xi=0$ \cite{Resnick_1986}, which is an exponential distribution.

Using eq.~\eqref{finite_mixture} we can write the finite exponential mixture density as:
\begin{equation}
f_k(x|\vec{\beta}, \vec{w}) = \sum_{i=1}^k w_i \beta_i e^{-\beta_i x} \, ,
\end{equation}
where $\vec\beta$ are the rate parameters of the exponential distributions with $\beta_i \in 
[0,\infty)$. To enforce the condition that the weights $\vec{w}$ sum to 1, the parameters are 
constructed from a set of unconstrained parameters $\vec{v}$ with $k-1$ degrees of freedom 
using the softmax function:
\begin{equation}
\label{softmax}
w_i = \frac{e^{v_i}}{\sum_{j=1}^{k}e^{v_j}} \, ,
\end{equation}
where $v_1=0$ and $v_{i\neq 1} \in (-\infty,\infty)$.

\section{Examples} \label{sec:examples}

The examples in this section are selected from public datasets available on \href{https://www.hepdata.net/}{\texttt{HEPData}} which is an online repository for publication-related high energy physics data \cite{Maguire_2017}. We select both large and small datasets corresponding to two jet and two photon final states to test the robustness of the method to dataset size and background composition.

In the following sections, we compare the finite exponential mixture to the original methods. The model parameters are estimated via maximum likelihood using the Python interfaces to \texttt{RooFit}~\cite{RooFit} and \texttt{ROOT}~\cite{ROOT}, which are commonly used in high-energy physics.
Numerical optimization is performed with \texttt{MINUIT} and the \texttt{MIGRAD} 
algorithm~\cite{Minuit2}.
Because the mixture likelihood is non-convex, multiple random restarts are used for each fit, with a limited number of retries when the minimizer stops in a flat region.

For each dataset, a sequence of models with increasing numbers of exponential components is fit, and the preferred value of $k$ is chosen by the Akaike information criterion (AIC), while also recording the Bayesian information criterion (BIC)
\footnote{The AIC and BIC are the penalized likelihood criteria $\mathrm{AIC}=-2\ell_{\max}+2p$ 
and $\mathrm{BIC}=-2\ell_{\max}+p\log n$, where $\ell_{\max}$ is the maximized log-likelihood, $p$ is the number of free parameters, and $n$ is the sample size. Both balance the goodness of fit against model complexity, but BIC penalizes additional parameters more strongly when $n$ is 
large and therefore tends to prefer simpler models.}, following the discussion of the mixture-model 
selection in McLachlan and Peel~\cite{finite-mixture-models}. The justification for preferring AIC in these studies is deferred to Sections~\ref{sec:toys} and~\ref{sec:discussion}. 
For ease of implementation, the rates are standardized relative to the single-exponential 
estimate $1/(\mathbb{E}(X)-X_{\min})$, and each rate is capped at 50 times that reference value to prevent overfitting at the lower 
boundary.

In \texttt{RooFit}, the normalization of the density depends on the defined observable range, which has consequences for model selection. Unless noted otherwise, we take $X_{\max}=10$~TeV, below the 13~TeV center-of-mass energy.
The remaining dataset-specific choices, including whether the likelihood is binned or unbinned and the exact restart settings, are given in the corresponding subsections.

\subsection{ATLAS Dijet Spectrum} \label{sec:example1}

The dataset used in this section was collected by the ATLAS experiment at the LHC during the 
Run~2 data-taking period, spanning from 2015 to 2018 .
The publication, Ref. \cite{ATLAS_dijet}, uses several signal regions defined by the half rapidity separation between the two leading jets $y^*= (y_1 - y_2)/2$, where $y_1$ and $y_2$ are the rapidities of the leading and subleading jets. For this example, we use the signal region defined by $|y^*|<0.6$. 

The dijet function used in this analysis is given by:
\begin{equation}
f_D(x) = p_0(1-x)^{p_1}/x^{p_2+p_3\log x},
\end{equation}
where $x=m_{jj}/\sqrt{s}$ for the dijet invariant mass $m_{jj}$ and center-of-mass energy 
$\sqrt{s}=13$~TeV.
Simpler versions of this function have been used in the past with $p_3=0$, but as datasets grow, these parameterizations must be extended. Because of this concern, ATLAS uses a sliding window fit method \cite{ATLAT_Dijet_sliding_window}, which reduces the fit range to the vicinity of the signal hypothesis being tested. For simplicity, we do not replicate this technique; instead fitting the entire spectrum at once.

With the exception of the $x^{p_3\log x}$ term, the components of the dijet function correspond to distinct physical phenomena: the finite energy threshold ($(1-x)^{p_1}$) and approximate scale invariance ($x^{p_2}$). At sufficiently high energies, Standard Model quarks can be treated as massless, yielding a theory that is approximately scale-invariant up to the running of the strong coupling constant $\alpha_s$. While a finite exponential mixture is not inherently scale-invariant, it can approximate scale-invariant distributions like $x^{p_2}$ (for $p_2<0$) arbitrarily closely as the number of mixture components increases. Furthermore, although the exponential mixture does not explicitly model the finite energy threshold, it can still provide a reliable approximation depending on where the upper tail is truncated.

The dataset available on \texttt{HEPData} is binned. In high-energy physics, datasets can be very large, making fits to the unbinned data time-consuming. In such cases, it is common practice to bin the data finely enough to retain information, while coarsely enough that the fits converge in reasonable times.
In this analysis the bin widths are determined by the dijet invariant mass resolution, which broadens at higher 
values.
The binned likelihood is then given by the product of Poisson probabilities:
\begin{equation}
    L(\beta) = \prod_b \text{Pois}(n_b|n_{f(\beta),b}),
\end{equation}
where $b$ is the bin index, $n_b$ is the observed count in bin $b$, and $n_{f,b}$ is the 
expected number of counts in bin $b$, given by:
\begin{equation}
    n_{f,b} = \int_{b_-}^{b_+} f(m|\beta)\,dm,
\end{equation}
where $b_-$ and $b_+$ are the lower and upper edges of bin $b$. Crucially, because the finite exponential mixture is a linear combination of exponential functions, this integral evaluates analytically to a simple closed form. By providing the exact integral to the minimizer, we eliminate the need for computationally intensive numerical integration routines, ensuring that the binned likelihood evaluation is performant and numerically stable during minimization.

The binning provided on \texttt{HEPData} includes only a single small-width bin following the 
last bin with non-zero entries.
Because model selection depends on the upper threshold of the fit, an additional bin with zero 
entries extending to 10~TeV is appended to constrain upper tail behavior.
Since this dataset is large, it is expected to be require a large number of parameters to model it 
adequately, implying more local minima in the likelihood surface.
Accordingly, 80 random restarts per model are used, with 100 retries per fit.

\begin{figure}
    \centering
    \includegraphics[width=\linewidth]{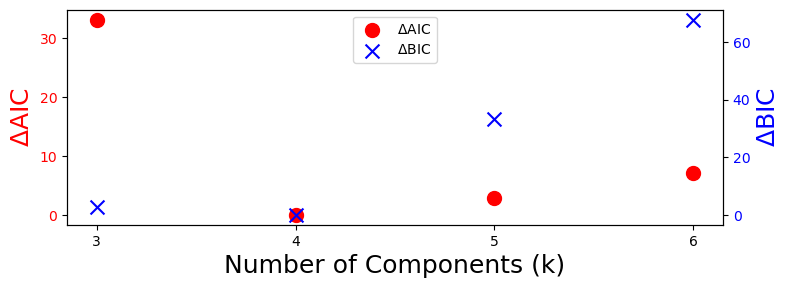}
    \caption{Relative AIC (red dot) and BIC (blue x) scores for exponential mixtures on the ATLAS dijet 
    dataset for $k=3,4,5,6$ mixture components. The relative AIC/BIC is given by $\Delta\text{IC}_i=\text{IC}_i-\min(\text{IC})$ where $\text{IC}$ is either AIC or BIC.}
    \label{fig:dijet_AIC_BIC}
\end{figure}

\begin{figure}[h]
    \centering
    \includegraphics[width=\linewidth]{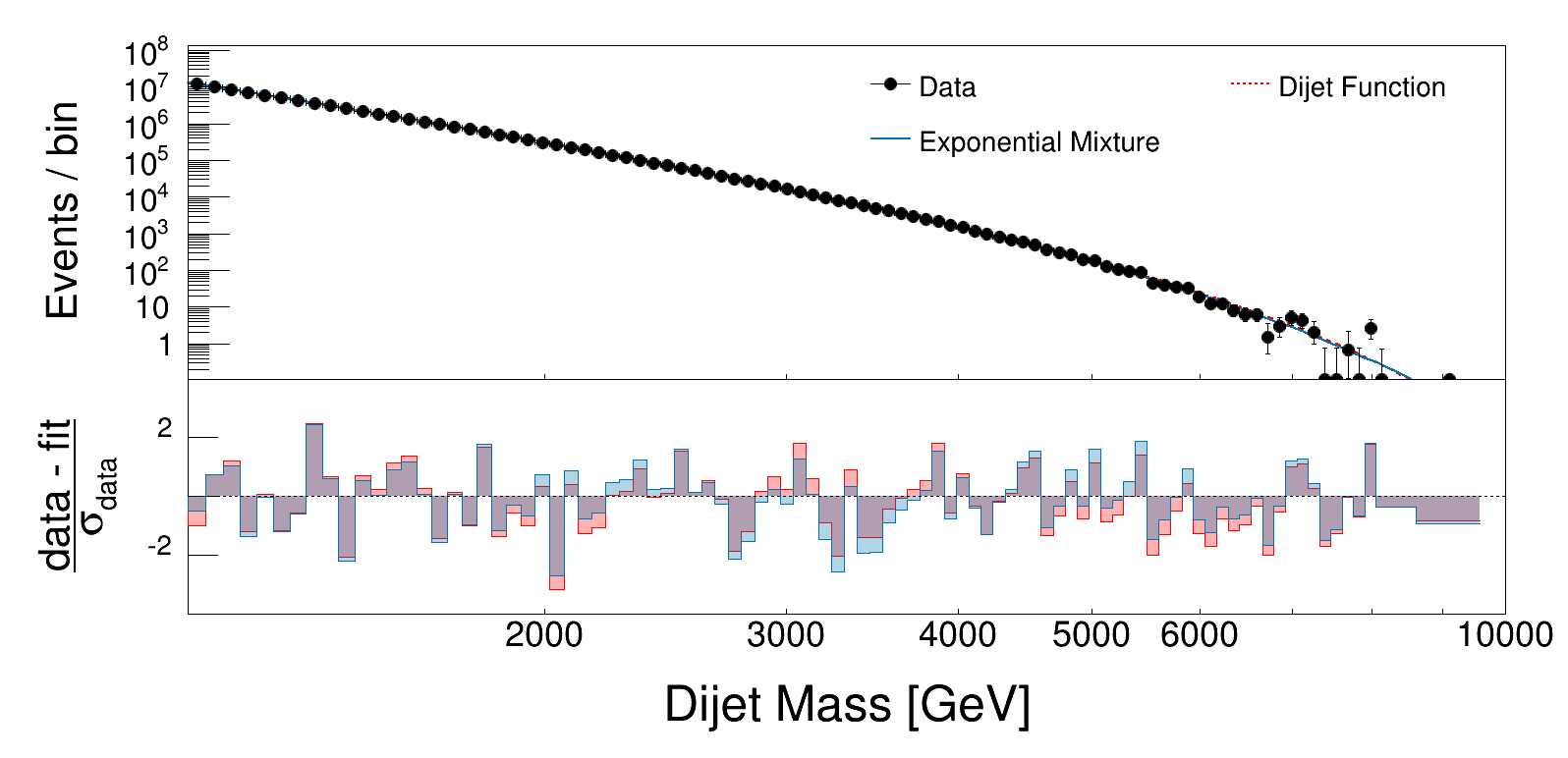}
    \caption{Fit of the dijet function and exponential mixture to the ATLAS Run~2 dijet dataset 
    (top) and normalized residuals (bottom).}
    \label{fig:dijet_fit}
\end{figure}

\begin{table}[h]
\centering
\begin{tabular}{lcccc}
\hline
Model & $\chi^2$ & $\chi_\nu^2$ & $\Delta$AIC & $\Delta$BIC \\
\hline
Dijet Function & 96 & 1.3 & 0.3 & 0 \\
Exponential Mixture & 90 & 1.3 & 0 & 60.4 \\
\hline
\end{tabular}
\caption{Goodness-of-fit metrics for the dijet function and finite exponential mixture (chosen by AIC) fit to the 
Run~2 ATLAS dijet data. For the $\chi^2$ calculation, low-event-count bins are merged until they contain at least 30 events.}
\label{tab:dijet_AIC_BIC}
\end{table}

Figure~\ref{fig:dijet_AIC_BIC} shows the AIC and BIC scores for various $k$; in this case, AIC and BIC agree, with the best model having four components.
The goodness-of-fit metrics are shown in Table~\ref{tab:dijet_AIC_BIC}: the exponential mixture achieves a slightly smaller $\chi^2$ but uses more parameters.
Figure~\ref{fig:dijet_fit} shows the fit of the dijet function and the 4-component exponential mixture to the dijet dataset.
The residuals show that the two functions provide similar estimates throughout the mass range.

\subsection{CMS High-Mass Diphoton Spectrum} \label{sec:example2}

The dataset considered in this section is the Run~2 dataset published by CMS in the high-mass two-photon final state \cite{CMS_high_mass_diphoton, diphoton_HEPDATA}.
Only the barrel--barrel sample (EBEB, $n=5,036$) is used here.

The background estimate for this analysis uses the discrete profiling method with the following 
functions:
\begin{align} \label{eq:diphoton_fns}
    f_1(x)&=p_0x^{p_1+p_2\log(x)}, \\
    f_2(x)&=p_0e^{p_1x}x^{p_2}, \\
    f_3(x)&=p_0(1+xp_1)^{p_2}, \\
    f_4(x)&=p_0(1+xp_1)^{p_2+p_3x},
\end{align}
where $x\equiv m_{\gamma\gamma}/\sqrt{s}$, $m_{\gamma\gamma}$ is the invariant mass of the 
two-photon system, and $\sqrt{s}=13$~TeV is the center-of-mass energy.
These functions are chosen from extendable classes of functions built using power laws, exponentials, and threshold-shifted polynomials, most of which are also used in the dijet function. For model specification, the primary analysis references \cite{CMS_DP1}, where the explicit criterion is to choose the lowest-order function yielding a $\chi^2$ goodness-of-fit $p$-value greater than 5\% for each functional family.

With the published parameter estimates on \texttt{HEPData} it can be shown that $f_2$ and $f_3$ are completely monotone functions.
The first function $f_1$ has completely monotone components, but the term $x^{p_2 \log(x)}$ is not; most departures from complete monotonicity are found at low $x$. The derivatives of $f_4$ are more complicated, with no guarantee of complete monotonicity, though it is possible for the published parameter estimates.

While the official \texttt{HEPData} entry provides a binned dataset, the unbinned data were 
obtained from the CMS collaboration upon request.
In this case, the likelihood is given by:
\begin{equation}
    \mathcal{L}(\vec\theta) = \prod_{i=1}^N p(x_i|\vec\theta),
\end{equation}
where $\vec\theta$ are the model parameters and $p$ is the probability density function. 
Given the relatively small sample size, 20 random restarts with 5 retries each are used. 

\begin{figure}
    \centering
    \includegraphics[width=\linewidth]{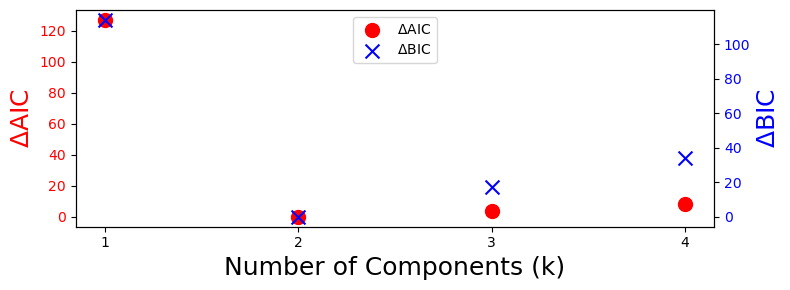}
    \caption{AIC (red dot) and BIC (blue x) scores for exponential mixtures on the CMS 
    high-mass diphoton dataset for $k=1,2,3,4$ mixture components. The relative AIC/BIC is given by $\Delta\text{IC}_i=\text{IC}_i-\min(\text{IC})$ where $\text{IC}$ is either AIC or BIC.}
    \label{fig:diphoton_model_selection}
\end{figure}

\begin{figure}[h]
    \centering
    \includegraphics[width=\linewidth]{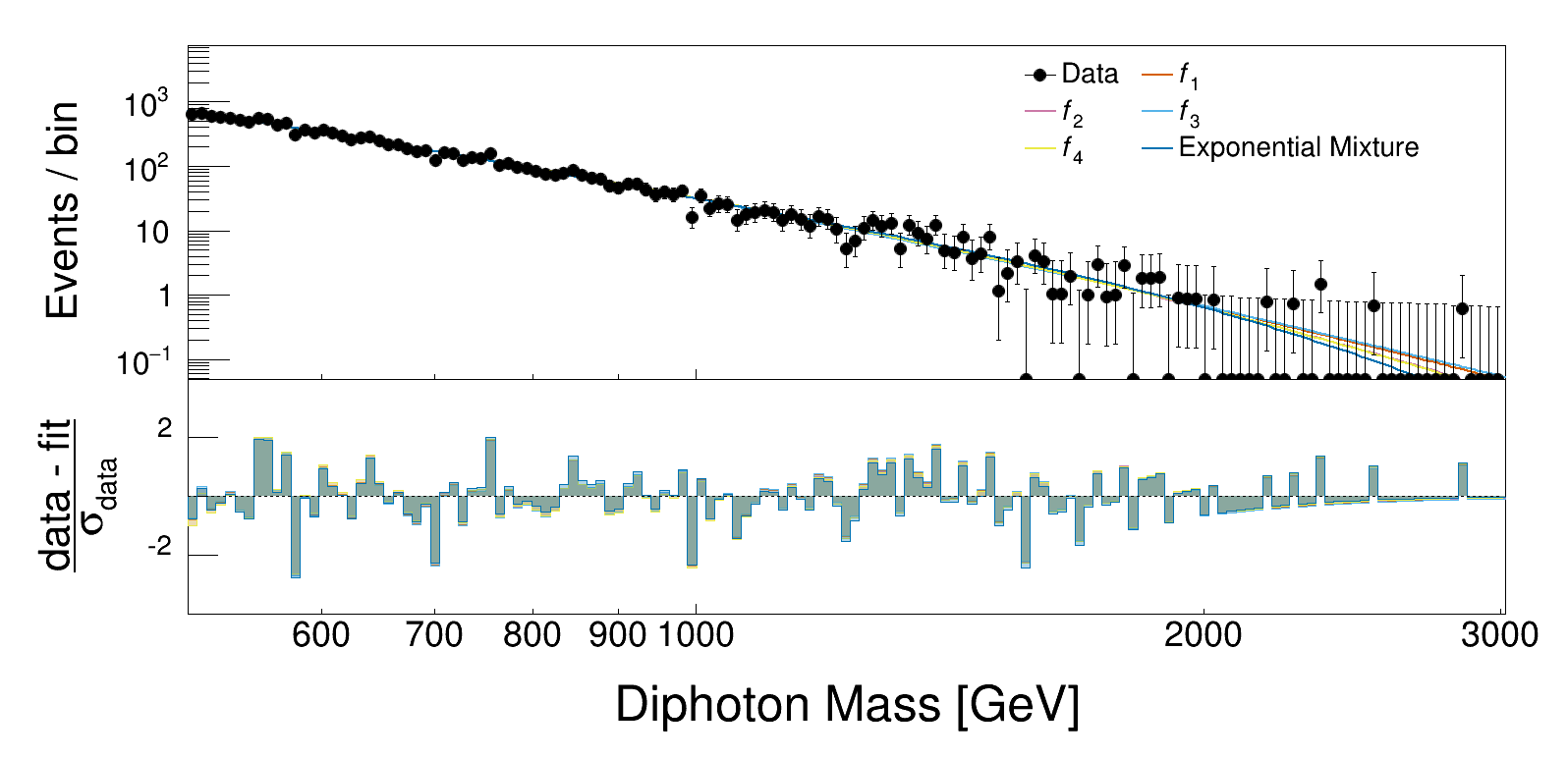}
    \caption{Fit of the original functions and exponential mixture (AIC) to the CMS high-mass diphoton dataset (top) and normalized residuals (bottom). For visualization, the data and density functions are binned.}
    \label{fig:diphoton_fit}
\end{figure}

\begin{table}[h]
\centering
\begin{tabular}{lcccc}
\hline
Model & $\chi^2$ & $\chi^2_\nu$ & $\Delta$AIC & $\Delta$BIC \\
\hline
$f_1$ & 54.2 & 0.93 & 1.6 & 1.6 \\
$f_2$ & 53.5 & 0.92 & 0.0 & 0.0 \\
$f_3$ & 54.7 & 0.94 & 2.5 & 2.5 \\
$f_4$ & 53.4 & 0.94 & 1.9 & 8.4 \\
Exponential Mixture & 50.6 & 0.90 & 1.2 & 14.2 \\
\hline
\end{tabular}
\caption{Goodness-of-fit metrics for $f_1$, $f_2$, $f_3$, $f_4$, and the finite exponential mixture ($k=2$) 
fit to the Run~2 CMS high-mass diphoton data. For the $\chi^2$ calculation, low-event-count 
bins are merged until they contain at least 30 events.} 
\label{tab:diphoton_AIC_BIC} 
\end{table}

Figure~\ref{fig:diphoton_model_selection} shows that both AIC and BIC prefer a two-component 
exponential mixture.
Table~\ref{tab:diphoton_AIC_BIC} shows that the exponential mixture fit quality is comparable 
to that of the original functions, with a slightly smaller $\chi^2$.
The fits of all four original functions and the exponential mixture are shown in 
Figure~\ref{fig:diphoton_fit}; the dataset is binned for visualization.

\section{Simulation Studies}\label{sec:toys}

In this section, the performance of the exponential mixture is quantified by fitting 
pseudo-datasets generated from each of the functions in eq.~\eqref{eq:diphoton_fns}. 
In each case, the true parameters are obtained via maximum likelihood estimation (MLE) on the 
CMS high-mass diphoton dataset.
The number of exponential components is determined independently for each pseudo-dataset, and results are shown for both AIC and BIC model selection metrics.
Agreement is quantified in terms of bias, coverage, and spurious signal yield.

\subsection*{Bias}
The bias here is measured as the average density difference scaled by the standard deviation of the 
density estimated using the true model. This relative bias $B(x)$, is given 
by:
\begin{equation}
    B(x) = \frac{1}{T\sigma(x)}\sum_{t=1}^T (f(x|\hat{\bm{\beta}}_t)-f_{true}(x))
\end{equation}
where $f_{true}$ is the true density used to generate pseudo-dataset $t$, $\sigma(x)$ is the standard deviation of the estimated densities given the true model, $f$ is the model being tested, and $f(x|\hat{\bm{\beta}}_t)$ is the MLE of $f$ for pseudo-dataset $t$. 

\begin{figure}[h!]
\includegraphics[width=\textwidth]{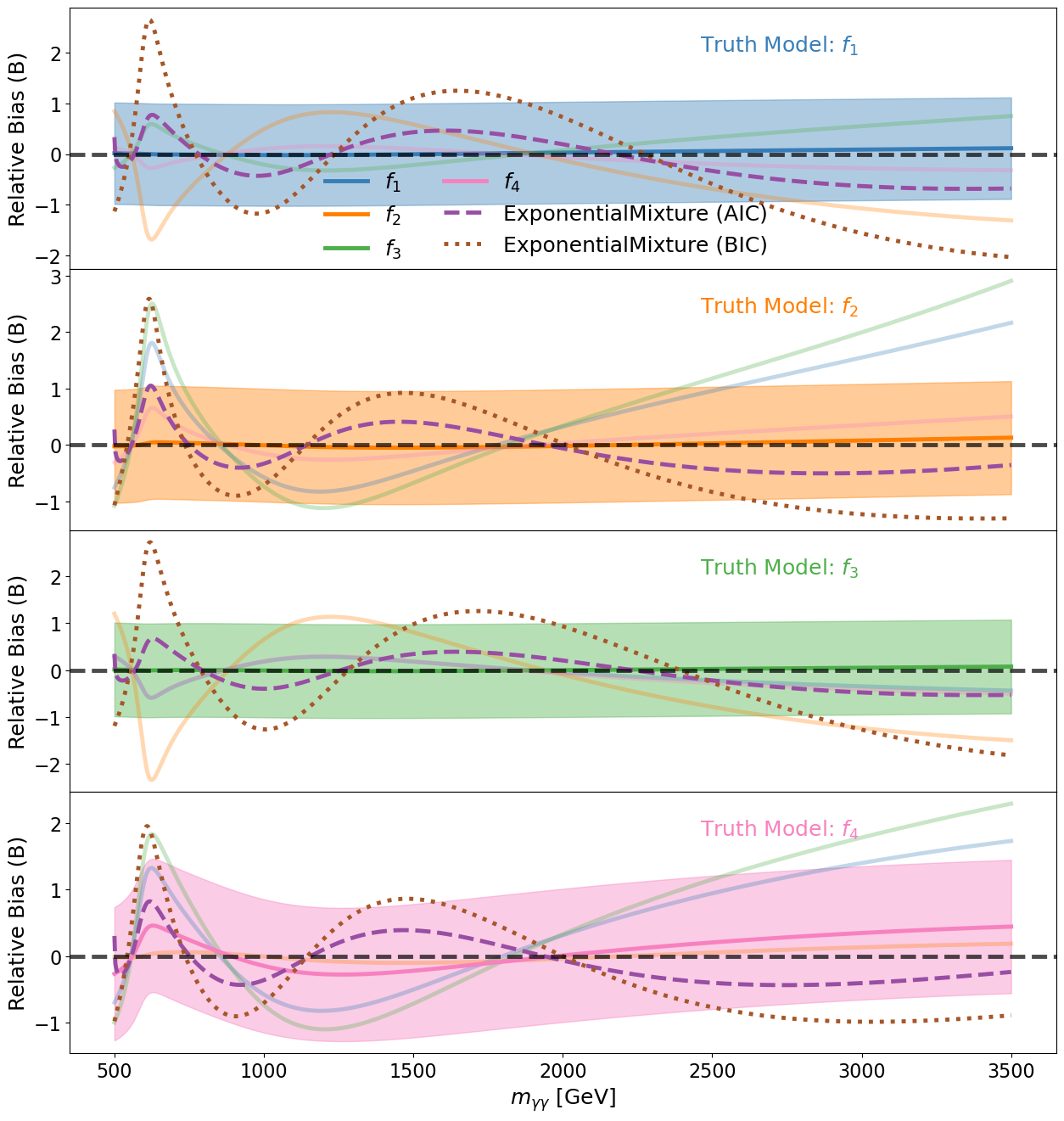}
\caption{Relative bias ($B(x)$) of the estimated density for various functions. Each row represents 
approximately $T=5000$ pseudo-datasets simulated from the true model with $n=5{,}036$ entries. 
The shaded regions represent the $\pm1$ standard deviation band for the true model estimates.}
\label{fig:bias}
\end{figure}

Figure~\ref{fig:bias} shows the relative bias of the exponential mixture with respect to
each of the four data-generating functions in eq.~\eqref{eq:diphoton_fns}, with fits to the remaining three shown for scale. 
These plots are assessed by their minima, maxima, and zero-crossings: a good bias function 
crosses the true value frequently and has small extremes.
The AIC-selected exponential mixture has the greatest number of zero-crossings and a bias 
contained mostly within the one-standard-deviation band of the true model, outperforming the 
BIC-selected mixture.
The AIC-selected mixture also achieves among the smallest biases across the competing models, though it uses more parameters than most.
In the tails, the bias remains within the uncertainty of the true model and decreases at high mass.

\subsection*{Coverage}

Coverage is evaluated by calculating how often the true density falls within the estimated uncertainty bands across the simulations.
Since fits to exponential mixtures of varying complexity may have weakly identified parameters, the likelihood surface can become irregular due to label switching, causing Hessian- or profile-based uncertainty estimates to be unstable.
To avoid these issues, the non-parametric bootstrap \cite{Efron_1986} is used to estimate 
confidence intervals.

A collection of $B$ bootstrap resamples $\{D_1,D_2,\ldots,D_B\}$ is drawn with replacement 
from the original dataset $D$, and the density is re-estimated on each resample to produce 
bootstrap replicas $\{\hat{f}(x|D_b)\}_{b=1}^B$. The empirical $(1-\alpha)$ confidence 
interval is then constructed from these replicas.

The coverage of model $f$ for true density $f_t$ is estimated as: 
\begin{equation}
    C(x|f,f_t,T,B) = \frac{1}{T}\sum_{i=1}^T I(x|D_i)
\end{equation}
where $T$ is the number of pseudo-datasets ($n=5{,}036$), $B$ is the number of bootstrap resamples, and 
$I(x|D)$ is the indicator function:
\begin{equation}
    I(x|D) =
    \begin{cases}
        1, & \text{if } f_{t}(x) \in \mathrm{CI}_B(x|f,D), \\
        0, & \text{otherwise,}  
    \end{cases}
\end{equation}
where $\mathrm{CI}_B(x)$ is the bootstrap confidence interval with $B$ resamples.

\begin{figure}[h!]
\includegraphics[width=\textwidth]{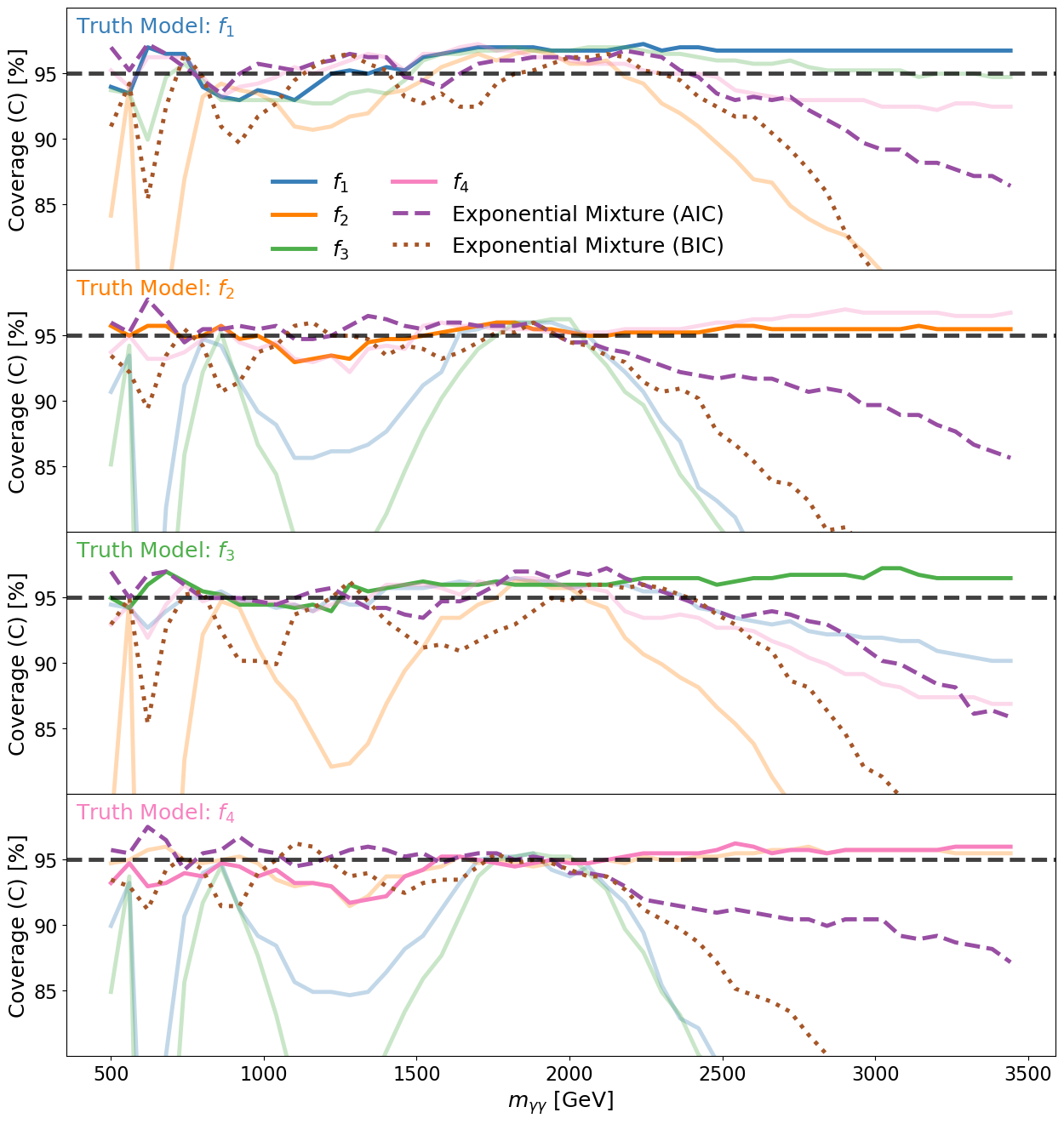}
\caption{Coverage of the estimated density for various functions. Each row represents $T=400$ 
pseudo-datasets simulated from the true model with $n=5{,}036$ entries. The 95\% confidence 
interval is calculated using $B=1000$ bootstrap resamples.}
\label{fig:coverage}
\end{figure}

Figure~\ref{fig:coverage} shows the coverage results.
As in the bias study, AIC outperforms BIC. The AIC-selected exponential mixture maintains nominal coverage across all data-generating models for $m_{\gamma\gamma}<2000$~GeV. In the tails, the AIC-selected mixture under-covers but nonetheless maintains one of the best 
coverages among the functions considered.

\subsection*{Spurious Signal}

The spurious signal yield of the data-generating models is compared to that of the exponential 
mixtures by fitting pseudo-datasets with a signal-plus-background model: 
\begin{equation}
    f(x)=N_b b(x)+N_s s(x),
\end{equation}
where $N_s$ is the signal yield, $s(x)$ is the signal model, $N_b$ is the number of background 
events, and $b(x)$ is the background model. The signal model is a Normal distribution with a standard deviation ranging from 30 GeV at the lowest mass hypothesis (600 GeV) to 60 GeV at the highest (2000 GeV). Signal yields are extracted over 1000 pseudo-datasets across this mass range.

\begin{figure}
    \centering
    \includegraphics[width=0.9\linewidth]{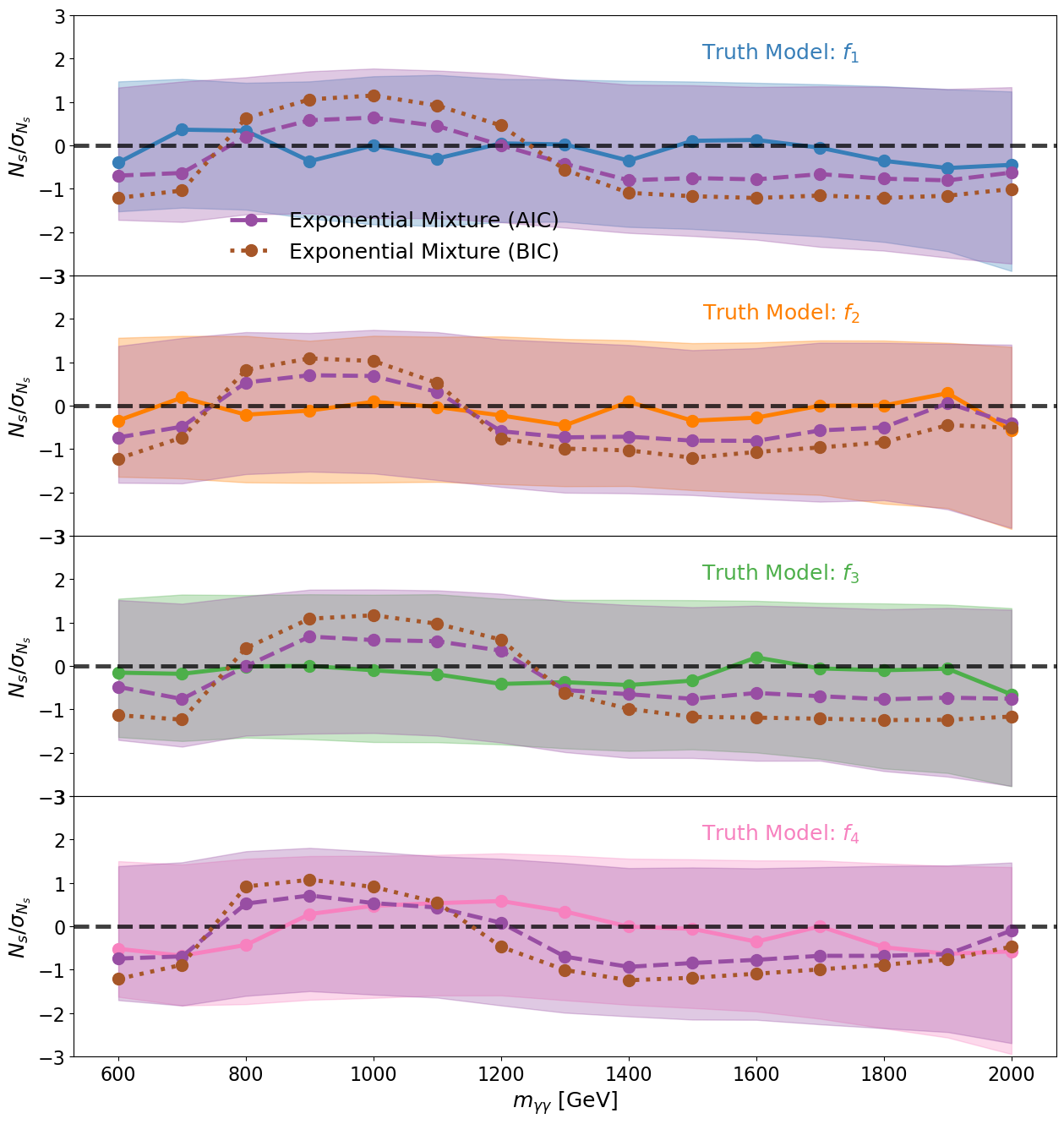}
    \caption{Expected signal yield $S(x)$ for the data-generating functions 
    and the exponential mixtures. Each point represents the median over $T=1000$ pseudo-datasets; bands represent the 68\% interval and are displayed only for the true model and the 
    AIC-selected exponential mixture.
    Each subplot corresponds to a different data-generating model.} 
    \label{fig:spurious}
\end{figure}

For each of the four data-generating functions in eq.~\eqref{eq:diphoton_fns}, the 
AIC- and BIC-selected exponential mixture signal yields are compared to those of the true 
models. For consistent scale, we plot the signal yield $N_s$ divided by the Hessian approximation of its uncertainty $\sigma_s$. Figure~\ref{fig:spurious} shows this quantity for several models and data-generating functions. We see that the median values of the AIC-selected exponential mixture are generally small compared to the variance in the true model, and the uncertainty in the exponential mixture largely covers zero. A negative value indicates that the exponential mixture over-predicts the density in that regio;, the range is restricted to 2000 GeV to prevent underflow that can happen in regions of low density with finite statistics. The spurious signal measured here is broadly consistent with the bias shown in Figure~\ref{fig:bias}: regions where the density is under-predicted correspond to over-prediction of the signal strength, and vice versa.

\section{Discussion} \label{sec:discussion}

The present work introduces the exponential mixture class as a flexible, asymptotically motivated framework for modeling smoothly falling backgrounds in high-energy physics. By prioritizing a basis of functions that satisfy intrinsic physical or Extreme Value Theory (EVT) constraints over rigid, analysis-specific functional forms, this approach hopes to offer a more generalizable alternative to traditional background modeling. Our results demonstrate that the exponential mixture class is broad enough to match the performance of existing methods across datasets that differ substantially in size and background complexity, as seen in the ATLAS dijet and CMS diphoton examples.

Standard background functions, such as the widely used dijet function, rely on theoretical approximations that are time-consuming to develop and which require revision as data volumes grow. Furthermore, complex final states involving photons or electrons may complicate the theoretical landscape. The exponential mixture is intended as a flexible, well-motivated alternative that reduces the need for exhaustive, case-by-case theoretical development.

While energy-based distributions are expected to be monotone in the absence of SM resonances due; complete monotonicity\footref{cm_def}, a constraint of the exponential mixture, is a stronger condition.
Several effects are strictly not completely monotone: the turn-on behavior near the lower 
threshold and the finite-energy cutoff at the upper end of the spectrum.
Analyses typically mitigate the turn-on behavior by imposing a sufficiently high lower 
threshold; alternatively, this behavior can be modeled independently and combined with the 
exponential mixture to capture the full spectrum. 

In addition, the exponential mixture will not model the finite upper bound. In the dijet example the data exhibit heavy tail behavior in the bulk but must eventually transition to a GPD $\xi<0$ near the upper bound. In transition, the shape may be approximately exponential, so a suitably truncated exponential 
mixture may still provide a reasonable approximation.
A more complete model could combine the exponential mixture with additional terms capturing 
behavior near the upper limit. For instance, one could extend the traditional dijet function by keeping the $(1-m_{jj}/\sqrt{s})^{p_1}$ term to model the finite-energy cutoff (acting as a GPD with a fixed upper endpoint at $\sqrt{s}=13$ TeV) while generalizing the power-law component with the exponential mixture.

When evaluating these mixtures, our framework shares similarities with methods where researchers utilize an extendable class of functions and rely on model selection to determine the ideal parameter count. When additional parameters have direct physical meaning, conservative test-based methods such as the likelihood ratio test can be more appropriate. Here, however, the simulation studies (with $n=5036$) show that the metric that prefers more parameters performs better. In particular, AIC selects larger mixtures than BIC, and those fits have more zero-crossings in the bias curves, smaller bias, better coverage, and smaller spurious signal. A natural explanation is that, in contrast to flexible models like polynomials, the exponential mixture is sufficiently rigid that additional components primarily reduce underfitting rather than introducing overfitting. In this regime, where overfitting is well controlled, model selection metrics that are more liberal than AIC may yield further improvements.

The exponential mixture is generally rigid enough to avoid most overfitting, though some 
regions require additional care. Near the lower threshold in particular, the mixture can fit 
local effects by introducing steeply falling components with small weights.
To mitigate this, the rates are parameterized relative to the single-exponential estimate 
$1/(\mathbb{E}(X)-X_{\min})$, with an upper bound of 50 times this reference rate; this choice 
reduces the risk of severe overfitting, though a more principled regularization criterion 
remains desirable.
Model selection is also sensitive to the upper threshold of the fit, a consequence of 
range-dependent normalization in \texttt{RooFit}. A substantial zero-event region above the 
last observation is included to constrain the fit in the tail, but the extent of this region 
remains an arbitrary choice that could be improved by more principled methods.

The principal practical challenges in this work concern model selection and optimization. 
The mixture likelihood surface is non-convex, with the number of local minima increasing with 
the number of components; false minima can also arise when component support points become too 
close together.
Additionally, as the number of components approaches or exceeds that required to model a 
dataset, the likelihood becomes very flat as some components acquire negligible weights. 
In the examples considered here, random restarts and retries were sufficient to find acceptable 
minima, but improved global optimization methods such as split-and-merge 
algorithms \cite{split-and-merge} warrant further study.

Finally, if the true distribution is completely monotone, the finite exponential mixture is guaranteed 
to converge to it as the number of components increases, but this does not imply that it is the most 
efficient basis at a finite sample size.
In the diphoton example, $f_2$ achieves a slightly smaller AIC than the exponential mixture 
(see Table~\ref{tab:diphoton_AIC_BIC}), and for the published parameter estimates $f_2$ is 
itself an exponential mixture.
This suggests that, even within the class of exponential mixtures, parameterizations or basis functions better suited to particular spectra may exist.

Overall, the exponential mixture is a credible, asymptotically and empirically motivated class of functions for modeling smoothly falling backgrounds, offering a reusable, constrained alternative to 
analysis-specific functional forms.
The principal open questions are concerning the optimal choice of complexity, robust optimization, and the incorporation of additional physics-motivated structure when the data require it.

\section{Conclusion} \label{sec:conclusion}

The finite exponential mixture has been considered as an alternative approach to modeling 
smoothly falling backgrounds in high-energy physics, motivated by extreme value theory.
This class provides a flexible semi-parametric basis for approximating falling distributions 
without committing to an analysis-specific functional form. In the ATLAS dijet and CMS 
high-mass diphoton examples, the exponential mixture gives fits comparable to existing methods despite substantial differences in dataset sizes and background complexity.
In pseudo-dataset studies based on the functions used by the published diphoton search \cite{CMS_high_mass_diphoton}, the AIC-selected exponential mixture has small bias, good coverage in the bulk, and small spurious signal.

These results suggest that the finite exponential mixture is a promising, statistically motivated general-purpose basis for modeling falling backgrounds in resonant searches. The model is broad enough to adapt to different, smoothly falling spectra while remaining rigid enough to avoid most overfitting; however, it exhibits clear limitations. Because the distributions of interest are not guaranteed to be completely monotone, non-completely-monotone effects—such as threshold turn-on behavior, finite-energy cutoffs, and other process-specific features—may need to be modeled separately. Furthermore, because performance depends on heuristics for model selection, fit range, and optimization, further work in these areas and in the treatment of non-completely-monotone effects will be essential to improving the performance and generality of this reusable model class.

\section*{Data Availability}
The dataset(s) analyzed during the current study are publicly available on \href{https://www.hepdata.net/}{HEPData}, see Ref. \cite{Maguire_2017} for more information. These data can be accessed at: \url{https://doi.org/10.17182/hepdata.91126} for the ATLAS dijet dataset and \url{https://doi.org/10.17182/hepdata.150677} for the CMS diphoton dataset.

\section*{Code Availability}
The custom code used to generate the results and figures in this manuscript is available on GitHub at: \url{https://github.com/atownse2/26.Townsend.expmix.unpublished}.


\bibliographystyle{JHEP}
\bibliography{biblio.bib}


\end{document}